\documentclass[aps,superscriptaddress,showpacs,floatfix,notitlepage,twocolumn]{revtex4-1}

\usepackage{amsmath}
\usepackage{amsfonts}
\usepackage{amssymb}
\usepackage{hyperref}
\usepackage{graphicx}
\usepackage{color}
\usepackage{xcolor}
\usepackage[sort&compress]{natbib}

\begin{document} 
\title{Photon correlations in both time and frequency}

\author{E. del Valle} 
\affiliation{
Departamento de F\'{i}sica Te\'{o}rica de la Materia Condensada, 
Universidad Aut\'{o}noma de Madrid, E-28049 Madrid, Spain}
\author{J. C. L\'opez Carre{\~n}o} 
\affiliation{
Departamento de F\'{i}sica Te\'{o}rica de la Materia Condensada, 
Universidad Aut\'{o}noma de Madrid, E-28049 Madrid, Spain}
\author{F. P. Laussy} 
\affiliation{
Faculty of Science and Engineering, University of Wolverhampton, Wulfruna St, Wolverhampton WV1 1LY, UK}
\affiliation{Russian Quantum Center, Novaya 100, 143025 Skolkovo, Moscow Region, Russia}

\date{\today}

\begin{abstract}
  While quantum mechanics precludes the perfect knowledge of so-called
  ``conjugate'' variables, such as time and frequency, we discuss the
  importance of compromising to retain a fair knowledge of their
  combined values.  In the case of light, we show how time and
  frequency photon correlations allow us to identify a new type of
  photon emission, which can be used to design a new type of quantum
  source where we can choose the distribution in time and energy of
  the emitted photons.
\end{abstract}

\maketitle

In quantum theory, one cannot have full information over a system.
This is one of the most important features of the theory, that caused
much exasperation to Einstein who insisted that in order to be
satisfactory, a physical model has to be
\emph{complete}~\cite{einstein35a}, i.e., describes all that can be
measured (what he called ``elements of reality'').  Let us take the
case of light. There are many things that one can measure from a light
field, such as its frequency (that corresponds to its color) or its
intensity (its brightness). In fact, one is typically interested in
the more complete information that provides the intensity at the
various frequencies, the so-called ``spectral distribution'' or
\emph{spectrum} of light. There is also the polarisation (if light
``points'' in a direction), the coherence (its ability to produce
fringes if superimposed to itself) and of course the position (in
space) where the field is measured or how these various quantities
change spatially.

\begin{figure}
  \includegraphics[width=.9\linewidth]{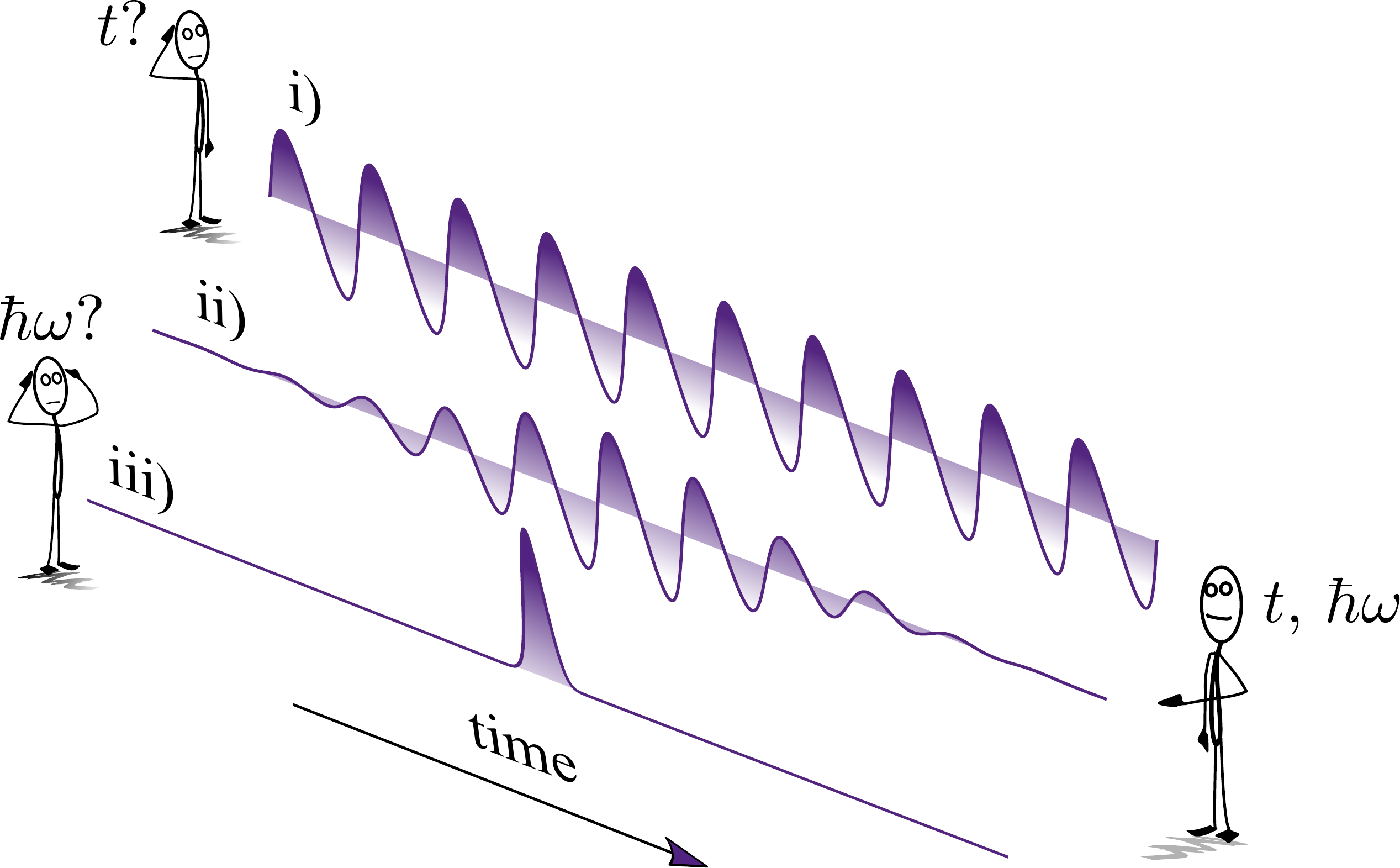}
  \caption{Three types of light-field propagation. One cannot know
    perfectly both the time of emission (or, equivalently, the
    position) of a photon \emph{and} its energy. This is because
    energy~$\hbar\omega$ in quantum mechanics is linked to the
    frequency~$\omega$ which, even in classical physics, is conjugate
    to time~$t$. The three cases shown are i) well-defined frequency,
    undetermined time, ii) compromise of simultaneously defined time
    and frequency and iii) well-defined time and unknown frequency.}
  \label{fig:Fri2Feb165754GMT2018}
\end{figure}

With the advent of quantum mechanics, our understanding of light
underwent considerable modifications, starting with a revival of one
of the great scientific controversies: the \emph{particle} versus
\emph{wave} nature of light. This had opposed Newton to Hooke,
Huygens, Young and other prestigious names but also to most the
evidence of the time. When light was later measured to travel slower
in a dense medium, the particle interpretation was believed to be
definitely buried. Two centuries later, Bose's derivation of the
\emph{blackbody radiation} (spectrum of emission due to temperature)
and the \emph{photoelectric effect} (how light produces current from
metals it shines on), related the energy, $E$, to the frequency,
$\omega$ (through Planck's relation $E=\hbar\omega$), and it appeared
that the light field is \emph{quantized}, that is, light is made up of
particles after all: the \emph{photons}. This exacerbated the
completeness problem as this makes particularly salient the
impossibility to know simultaneously two basic and crucial quantities
for a photon: its energy and position (or, equivalently, its time of
emission or time of detection, etc.) Time and frequency are indeed
\emph{conjugate}, meaning that they refer to complementary features
that cannot be defined together. This is illustrated in
Fig.~\ref{fig:Fri2Feb165754GMT2018} for three possible cases of the
light field: i) where the energy (frequency) is perfectly known,
resulting in complete indeterminacy in the time of emission, iii) the
opposite case where the time of emission is perfectly known, resulting
in indeterminacy of frequency and ii) a compromise where both energy
and time are known within some finite accuracy. Since all the evidence
in favour of the wave interpretation could not be disposed of, but
remained in startling contradiction, the view emerged of the
\emph{wave-particle duality}. This was one of the first uncanny
quantum-mechanical concepts, arguing for the coexistence of mutually
excluding aspects.

Light is even more bizarre when one considers its \emph{correlations}.
The concept of ``correlation'' describes the tendency of variables to
exhibit some degree of relationship, which can be small or even zero
(``uncorrelated'') or on the contrary large or even total (in which
case one can equate them through a mathematical function).  A shocking
property of light was discovered by Hanbury
Brown~\cite{hanburybrown56b} in the mid 1950s, with the observation
that the photons arrival times on a detector are positively
correlated---meaning that photons tend to arrive together, in
``bunches''---and this even if they come from different, uncorrelated
sources, such as two stars from separate galaxies. This always happens
as long as detectors are ``blind'' to the photonic properties (i.e.,
they do not resolve their frequency, polarisation, etc.) Such dramatic
and counter-intuitive properties caused a commotion and were even initially
rejected as absurd by the scientific community, but it was soon
understood~\cite{purcell56a} as a manifestation of
indistinguishability of particles which can be accumulated in the same
state (``bosons'' or wave-like objects), as opposed to those which
cannot co-exist (``fermions'' or matter-like objects).  This led to
the development of \emph{quantum optics}~\cite{glauber06a} and brought
another revolution in our understanding of a central theme of Physics:
\emph{coherence}. Rather than being linked to the monochromaticity of
a field and the stability of its amplitude, coherence is more
fundamentally a measure of the degree of correlations between photons.
A coherent light source, like a laser, is one for which photons have
no mutal correlations. This happens when the granularity of the field
does not matter and removing a photon leaves the field essentially
unperturbed, corresponding to the classical case where something can
be observed without affecting it. As a consequence, photons arrive at
the detector without any time relationship between them, as one would
have expected in the first place.  Blackbody radiation, where photons
are in thermal equilibrium, comes on the contrary with strong positive
correlations, namely, the bunching reported by Hanbury Brown. There is
a third, negative type of correlations, which is of special interest
as it describes quantum states of the light field which have no
classical counterpart and can power quantum technology. A single
photon is an example. Measuring the light field in this case removes
the photon and one is left with the vacuum.  Such states of the
light-field are highly sought after because they can be used for
so-called ``quantum information processing''~\cite{nielsen_book00a},
where ``qubits'' replace the bits (0 or 1) to enhance considerably the
computing power to the point that computation currently out of reach,
such as factoring very large integers, would become feasible. The case
of integer factorisation would result in breaking most protocols of
cryptography currently in use on the planet. To do so, one needs of
course $n$ correlated photons (for some integer~$n$) to be used as the
many qubits required by the quantum computer.

In the pursuit of such applications, quantum optics became largely the
science of time correlations between photons. This is measured by a
so-called $n$-th order correlation function $g^{(n)}(t_1,\cdots,t_n)$,
which is a joint probability density for the detection of photons at
times $t_i$, for $1\le i\le n$. The simplest case is that of two
photons emitted by a continuous source, where the time-delay betwen
photons, $\tau=t_2-t_1$, is what matters, so that one can deal with
the quantity $g^{(2)}(\tau)$. A typical quantum source is one for
which $g^{(2)}(0)=0$, meaning that photons do not arrive together, in
contrast to their bosonic tendency of bunching. Such a source can be
made for instance with a \emph{two-level system}, such as an atomic
transition or its solid-state counterpart, an ``artificial atom''
known as a ``quantum dot''. Such a system has only a ground and an
excited state, so when it makes \emph{transitions} (going from one
level to the other), it does so one-photon at a time.

Now, since it is impossible to have a complete characterization of
photons in both time and frequency, photon correlations have largely
been focused on their temporal aspect alone.  It is, however,
important for quantum applications that photons remain
indistinguishable, or the quantum effects would be washed out and a
stream of photons would reduce to a mere classical stream of
bits. This means that if frequency is to be resolved in a measurement,
photons must be closer in frequency than the detector resolution.  To
properly describe what happens when one compromises between time and
frequency, one needs a \emph{theory of photon correlations} not only
in time alone, but that covers for these two variables. The formal
aspects of such a theory have been developed in the late
80s~\cite{knoll86a, cresser87a, knoll90a} by providing the
mathematical expression for the joint probability distribution
$g^{(n)}_{\Gamma_1,\cdots,\Gamma_n}(\omega_1,t_1;\cdots;\omega_n,t_n)$
to detect $n$ photons such that the one detected at time $t_i$ has
energy~$\omega_i$. The uncertainty has been brought here in frequency
through the detector resolution (the so-called ``filter \emph{spectral
  linewidth}'')~$\Gamma_i$, that gives the accuracy one has on
$\omega_i$. Physically, this corresponds to filtering the light and
thus determining (or measuring) its frequency.

\begin{figure}
  \includegraphics[width=.95\linewidth]{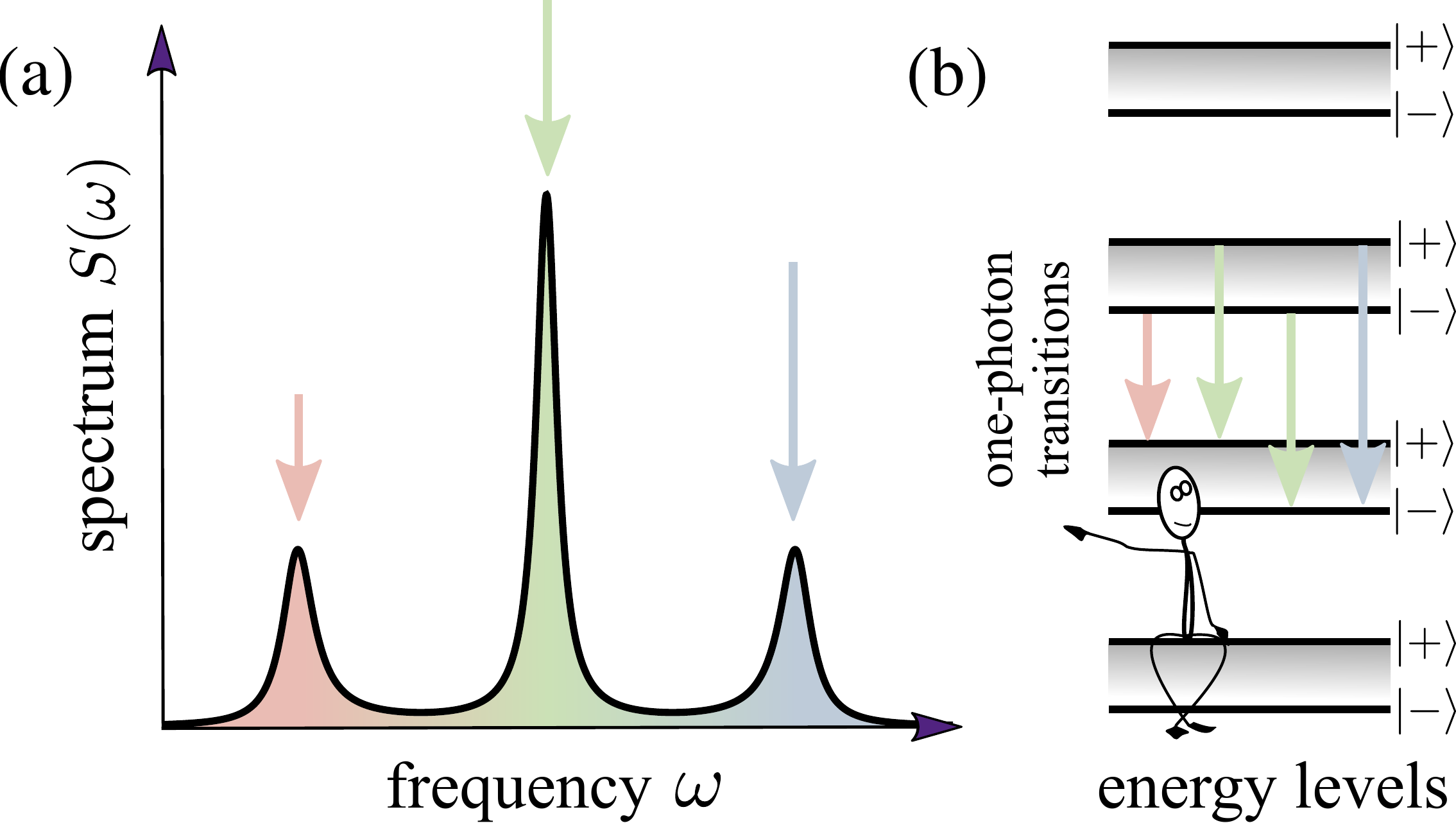}
  \caption{The spectrum of emission (distribution of emitted
    frequencies) of a coherently driven two-level system, known as
    resonance fluorescence, features three peaks (the so-called
    ``Mollow triplet''). This can be understood as photon transitions
    between neighbouring doublets of the level structure, shown to the
    right. As two of the four possible transitions have almost the
    same energy, one peak is twice as high as the others. A natural
    question that this structure pauses is: how are the photons from
    the various peaks correlated?}
  \label{fig:2}
\end{figure}

One fundamental source of light has been particularly fit for the
study of correlations with combined frequency \emph{and} time
information: \emph{resonance fluorescence}. This consists of a laser
exciting a two-level system, which emits light by spontaneous emission
(rather than by scattering or reflecting the laser; this gives the
``fluorescence'' part of the name) and when the energy of the laser
matches the energy of the system's transition (this gives the
``resonance'' part of the name). This is therefore a fundamental case
of driving a system at the frequency at which it emits. In such a
case, the spectrum of emission consists of a triplet, shown in
Fig.~2(a), that is called the \emph{Mollow triplet} after the person
who first provided its mathematical expression~\cite{mollow69a}.  The
physical reason for this peculiar lineshape was provided through the
so-called \emph{dressed-atom} picture~\cite{cohentannoudji77a}, that
shows how the combined two-level system plus resonant driving leads to
a level structure that consists of an infinite ladder of doublets,
shown in Fig.~2(b). The doublet comes from the two-level system and
its infinite repetition comes from the photons of the laser. Now,
photon transitions from one doublet to the next account for the
spectral structure in much the same way that photon transitions
between the states of the hydrogen atom account for its spectral lines
(Lyman, Balmer, Paschen and other series). The exact computation of
frequency and time-resolved correlations of this simple system,
however, remained out of reach even for the case of two photons only,
$g^{(2)}(\omega_1,t_1;\omega_2,t_2)$, until a technique was introduced
by some of the Authors~\cite{delvalle12a} that allows to compute this
quantity without the approximations perfomed before. In particular, it
lifted an important constrain of previous works to limit to photons
whose energies are those of the emission peaks. This may not seem to
be a serious limitation since the system emits mainly at these
frequencies.  However, computing the full two-photon correlation
spectrum (2PS) of resonance fluorescence~\cite{gonzaleztudela13a}, it
was found that, on the contrary, most of the interesting quantum
emission does arise \emph{away} from the peaks. The 2PS for resonance
fluorescence is shown in Fig.~3(a), showing the joint probability
distribution of detecting two photons at the same time for all the
possible combinations of frequencies. This results in a
two-dimensional landscape (as there are two frequencies). The color
code is such that red corresponds to bunched photons (increased
probability to detect two photons with the corresponding frequencies
together), white is for no correlations (same probabilities as for two
random sources) and blue is for antibunched photons (decreased
probability to detect the photons simultaneously). The point at (1,0),
for instance, corresponds to correlations between photons coming from
the central and high-energy satellite peaks. This is blueish, meaning
that such photons tend to avoid being detected together.

The computation of this complete landscape of photon correlations let
appear a clear feature: the three antidiagonal lines of strong
bunching. These correspond to so-called ``\emph{leapfrog}
transitions'', whereby transitions in the ladder of states of
Fig.~2(b) are not between neighbouring doublets anymore, but jumping
over one of these (hence the name) in a two-photon transition.  The
fact that the transition now occurs with two photons lifts the
quantization of the spectrum, since only the sum $\omega_1+\omega_2$
is quantized, so that $\omega_1$ can take any value, as long as the
other photon fulfils energy conservation by being emitted at
frequency~$\omega_2=(E/\hbar)-\omega_1$. This is actually a known phenomenon
from planetary nebulae that results in a rare situation where atoms
are trapped in states which only have a two-photon channel escape,
resulting in continuous spectra that have puzzled astrophysicists for
some time~\cite{spitzer51a}. In our case, there is no need of a lucky
suppression of single-photon events, these can remain dominant as is
the case in resonance fluorescence where most of the emission is
indeed coming from the peaks. The filtering in frequency allows to
literally unravel the two-photon emission, which one can show is
maximally nonclassical~\cite{sanchezmunoz14b}. This theoretical
prediction for the structure of photon correlations has been confirmed
experimentally with a spectacular agreement~\cite{peiris17a}.

\begin{figure}
  \includegraphics[width=\linewidth]{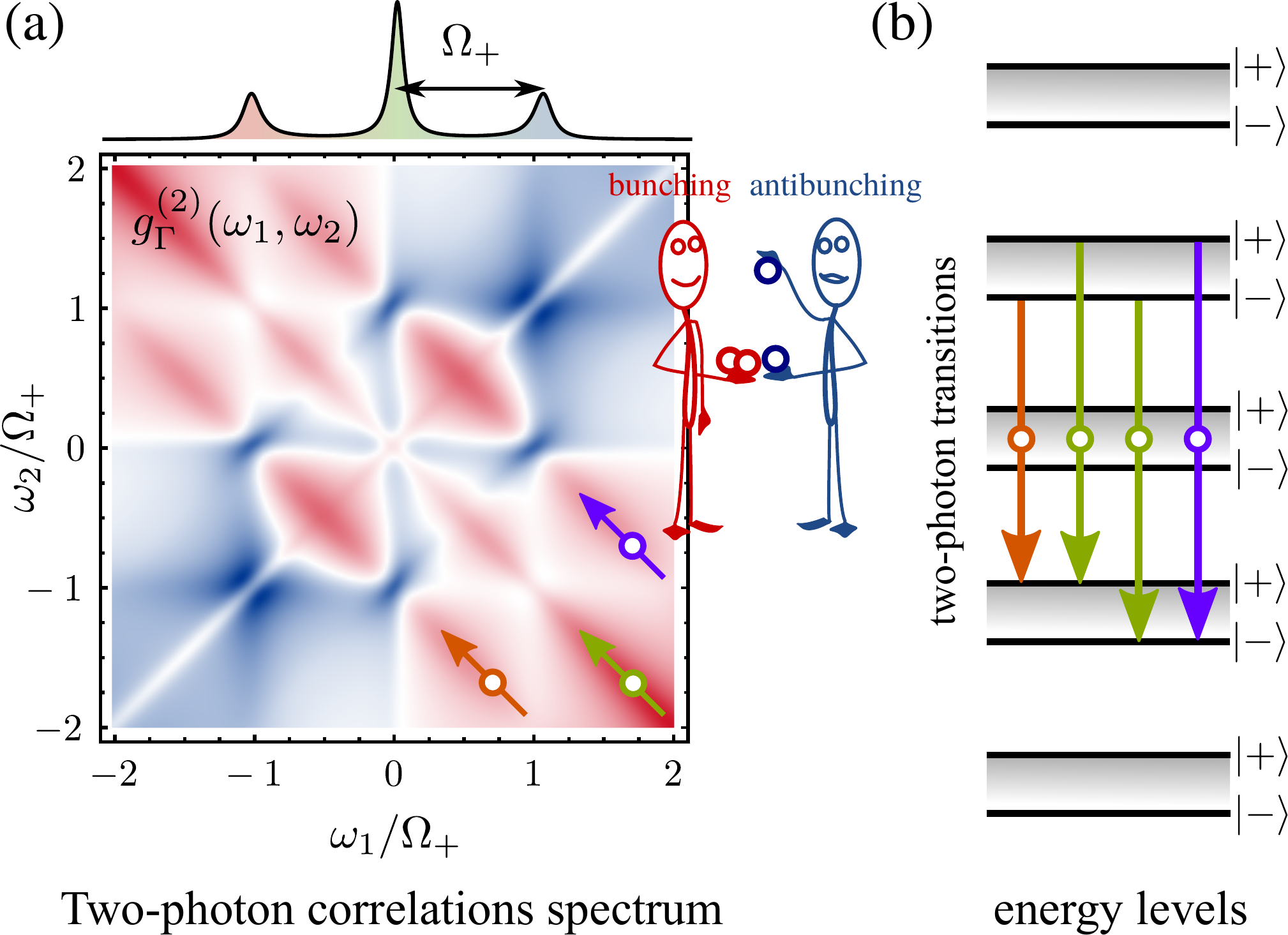}
  \caption{The two-photon correlation spectrum of the Mollow triplet,
    that shows $g_\Gamma^{(2)}(\omega_1,\omega_2)$ for all possible
    combinations of frequencies~$\omega_1$ and~$\omega_2$, in units
    of~$\Omega_+$ the splitting (distance) between the central and
    satellite peaks. The blue regions correspond to antibunching,
    meaning that photon pairs with the corresponding frequencies occur
    rarely. The red regions correspond to bunching, meaning that
    photon pairs with the corresponding frequencies are on the
    opposite more frequent than expected. White areas correspond to no
    correlations between the photons. The three red antidiagonal
    reveal a new type of processes: the ``\emph{leapfrog
      transitions}'' (shown right), that correspond to two-photon
    transitions jumping over an intermediate doublet of levels of the
    system. This happens far from the peak and is a source of
    correlated quantum light.}
  \label{fig:3}
\end{figure}

Computing photon correlations in both time and frequency thus led us
to the discovery of a new type of photon emission which had remained
unnoticed despite five decades of combined theoretical and
experimental scrutiny on the fundamental problem of resonance
fluorescence~\cite{cohentannoudji79a,aspect80a,ulhaq12a}. The
importance of this finding lies in the obvious prospects it enables
for the design of new types of photon sources. Quantum mechanics is
notorious for making it possible for anything to happen, by providing
a probability amplitude to any event whatsoever which, at the
classical level, may be small or cancelled by others with which it
interferes destructively. Resonance fluorescence is such a quantum
system that, although it is globally a single-photon source that stems
from a two-level system, actually embeds any type of photon emission,
which one can distillate and harverst through frequency
filtering~\cite{delvalle13a}. One can focus on any desired event by
coupling the system to a photonic resonator, known as a ``cavity''.
This stimulates the emission at the frequencies of interest by a
process known as ``Purcell-enhancement''.  For instance, placing the
system in a single cavity with a frequency at $1/n$th of the distance
(in frequency) between the central and satellite peak, turns the
Mollow triplet into a pure source of $n$-photon
emission~\cite{sanchezmunoz14a,sanchezmunoz18a}. That is to say, this
generalizes the case of a single-photon source to one that emits
exactly and exclusively $n$ photons, for any integer~$n$ (chosen by
placing the cavity at the adequate frequency). But even this already
remarkable extension only scratches the surface. By turning to more
complex types of leapfrog transitions~\cite{lopezcarreno17a}, one can
realise other types of versatile and tunable quantum sources. For
instance, one can design a configuration where a photon of a given
energy ``heralds'' (announces) the subsequent emission of five
photons equally split in frequency, that can be used as an input for a
quantum gate. The number of photons and their distributions in energy
are configurable, by placing cavities at the corresponding
frequencies. This may be technically challenging, but the principle is
simple and general enough to inspire actual implementations, in this
or other quantum systems. The ability to tune and exploit photon
correlations thus promises a wealth of applications, ranging from
quantum spectroscopy~\cite{lopezcarreno15a} to providing better
quantum sources of the types already known, or of a completely new
character~\cite{lopezcarreno16a,lopezcarreno16b}.

In conclusion, although quantum mechanics does not allow us to know
everything about a system, it is important to deal with the compromise
on knowledge one can get from all the conjugate variables.  In the
case of photon correlations, we have shown that the time-information
alone, which has dominated the field of quantum optics since its
creation, gives an overly simplified picture of the structure of
photon emission. This, in particular, has kept hidden important
leapfrog processes that jump over the states by involving several
photons at once. Such processes can power the quantum technology of
tomorrow, for instance by turning such a simple and fundamental
problem as resonance fluorescence into a configurable universal
multiple-photon emitter.

\bibliographystyle{naturemag}
\bibliography{Sci,books} 

\end{document}